\documentclass[twocolumn,floats,prl,aps,superscriptaddress,showpacs]{revtex4}

\usepackage[dvips]{graphicx}
\usepackage{latexsym}
\usepackage{euscript}

\begin{document}

\title{Transient Infrared Photoreflectance Study of Superconducting 
MgB$_2$: Evidence for Multiple Gaps and Interband Scattering}
       
\author{R.P.S.M. Lobo}
\email{lobo@espci.fr}
\affiliation{Laboratoire de Physique du Solide (UPR 5 CNRS) ESPCI, 
10 rue Vauquelin 75231 Paris, France.}

\author{J.J. Tu}
\affiliation{Department of Physics, The City College of New York, 
New York, NY 10031.}

\author{Eun-Mi Choi}
\author{Hyeong-Jin Kim}
\author{W.N. Kang}
\author{Sung-Ik Lee}
\affiliation{National Creative Research Initiative Center for 
Superconductivity, Department of Physics, Pohang University of 
Science and Technology, Pohang 790-784, Korea}

\author{G.L. Carr}
\affiliation{National Synchrotron Light Source, Brookhaven National 
Laboratory, Upton, NY 11973.}

\date{\today}

\begin{abstract}
We studied the far-infrared reflectance of superconducting MgB$_2$ 
in and out of equilibrium using pulsed synchrotron radiation. The 
equilibrium data can be described by a single energy gap at 
$2 \Delta /hc = 43 \textrm{ cm}^{-1}$. The fit improves slightly at 
lower frequencies when two gaps (at $30 \textrm{ cm}^{-1}$ and 
$56 \textrm{ cm}^{-1}$) are assumed to exist. The non equilibrium 
data is obtained when excess quasiparticles are generated in the 
MgB$_2$ film by a laser pulse, decreasing the energy gap(s). The 
temperature dependence of the number of excess quasiparticles 
indicates that only a single gap near $30 \textrm{ cm}^{-1}$ 
controls the non-equilibrium process. Fits to the photo-induced 
reflectance show the presence of the two gaps but weakening of an 
energy gap at $30 \textrm{ cm}^{-1}$ only. Our data can be understood 
if one considers the picture where excited quasiparticles created 
by the laser impulse relax to the lowest gap band before recombining 
into Cooper pairs. 
\end{abstract}


\maketitle 

The heat capacity of MgB$_2$ was known down to 20 K since 1957 but the 
sparse data set did not make the superconducting transition clear \cite{Swift}. 
More recently, detailed data revealed superconductivity in this compound with a
remarkably high critical temperature \cite{Nagamatsu} for a system showing a 
classical electron-phonon coupling mechanism. The calculated Fermi surface 
for MgB$_2$ has two separate pieces --- a 2D ($\sigma$) and a 3D ($\pi$) 
bands \cite{Kortus}. Liu {\it et al.} \cite{Liu} propose that two energy gaps 
($\Delta_\pi$ and $\Delta_\sigma$) exist for these two portions of the Fermi 
surface and predicted that $\Delta_\sigma \approx 3 \Delta_\pi$. This picture 
is supported by experiments such as tunneling \cite{Giubileo}, Raman 
spectroscopy \cite{Quilty}, heat capacity \cite{Bouquet} and angle-resolved 
photo-emission \cite{Souma}. Tunneling data also shows that in the 
superconducting state the two bands undergo the superconducting transition 
at the same temperature \cite{Giubileo}.

Two-band superconductivity is an interesting problem that has been addressed 
by Suhl {\it et al.} \cite{Suhl}. They showed that for 
non-interacting bands two critical temperatures exist. If a very small 
interaction between the bands is turned on, the whole system shows a 
superconducting transition at the highest $T_c $while preserving the two gaps. 
Recent calculations \cite{Choi} showed that this weakly interaction scenario 
applies to MgB$_2$. However the mechanism and size of the interaction remain 
speculative. Infrared spectroscopy on MgB$_2$ \cite{Jung,Tu,Pimenov,Kaindl1,Perucchi} did 
not provide any clear cut picture about the presence of two bands in MgB$_2$ 
although the data does not seem to follow strictly the BCS theory. 

In this letter we report an infrared reflectance study of a MgB$_2$ thin film, 
including the transient change in reflectance that results when a laser pulse 
produces excess quasiparticles in the film. We find that the conventional 
reflectance spectrum can be roughly described by an optical conductivity based 
on a single energy gap near $43 \textrm{ cm}^{-1}$. However, the agreement 
between theory and experiment improves at low frequencies if one assumes 
two energy gaps at 30 and $56 \textrm{ cm}^{-1}$. The transient 
photo-reflectance technique senses how the energy gap changes due to an excess 
population of quasiparticles. The temperature dependence of this excess 
quasiparticle signal obtained from the time dependent changes in the 
photo-reflectance measurements on the same film is fully described by a 
single energy gap at $30 \textrm{ cm}^{-1}$, in apparent contradiction with 
the conventional reflectance results. This discrepancy is resolved by the 
photo-reflectance spectrum that indicates the presence of at least two energy 
gaps in the system with the non-equilibrium dynamics dominated by the smaller 
energy gap. 

The sample for our study was a thin film of MgB$_2$ on a sapphire 
substrate \cite{Kang}. The film is estimated to be about $30 \textrm{ nm}$ thick 
and highly oriented with the c-axis normal to the substrate surface. Typical 
of many thin superconducting films, the $T_c$ of 30 K is suppressed compared 
to bulk material ($T_c = 39$ K) \cite{Jung,Tu}. Upon cooling, the 
electrical resistance decreases slightly, indicating metallic behavior 
dominated by impurity-type carrier scattering. Standard reflectance 
measurements were performed using the Bruker IFS 66v FTIR spectrometer at 
beamline U10A of the NSLS, with synchrotron radiation as the IR source. 
The specimen was solidly clamped with indium gaskets to the copper 
cold-finger of a heli-tran cryostat, leaving a 3 mm diameter aperture exposed 
for the IR measurement. The remaining 80\% of the sample's surface was 
available for thermal conduction into the cold finger. The far-infrared 
reflectance from the film was measured at a variety of temperatures below 
$T_c$ (superconducting state) and at $T = 35$ K (normal state). 

\begin{figure}
  \begin{center}
    \includegraphics[width=8cm]{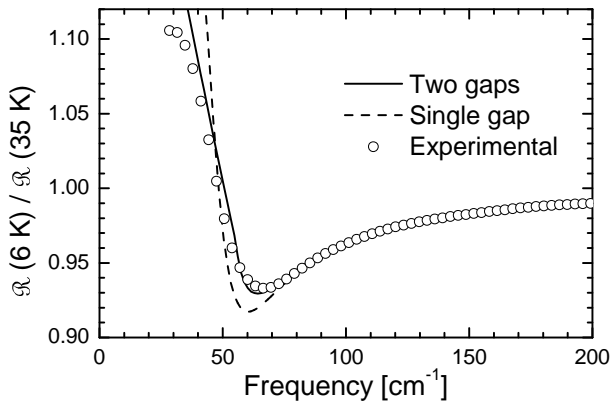} 
  \end{center}
\caption{Measured ratio of 
${\EuScript R}(T = 6\textrm{ K})/{\EuScript R}(T =35\textrm{ K})$ for a 
MgB$_2$ film (open circles). The lines are fits assuming either a single 
gap $2 \Delta = 43 \textrm{ cm}^{-1}$ (dashed curve) or a simple two gap 
model (solid curve) with $2 \Delta_\pi = 30 \textrm{ cm}^{-1}$ (55 \%) 
and $2 \Delta_\sigma = 56 \textrm{ cm}^{-1}$ (45\%). In both fits we used 
$R_\Box = 18~\Omega$; 
$\tau^{-1} = 300 \textrm{ cm}^{-1}$ and 3.05 as the refraction index for 
the sapphire substrate.} 
\label{fig1}
\end{figure}

The ratio of the normal state to the superconducting reflectivity at 
6 K, ${\EuScript R}_S/{\EuScript R}_N$, is shown in Fig. \ref{fig1} as open 
circles. The lines are fits to the reflectivity using an expression for the 
reflectance of a thin film on a transparent substrate in terms of the film's 
optical conductivity
\begin{equation}
  {\EuScript R} = \frac{(n-1+y_1)^2+y_2^2}{(n+1+y_1)^2+y_2^2} 
  \label{eq1}
\end{equation}
where $n$ is the substrate's refractive index and 
$y = (4 \ \pi / c) \hat{\sigma}d$ is the film's dimensionless complex 
admittance with $\hat{\sigma}$ the optical conductivity and $d$ the film 
thickness \cite{Gao}. This expression is valid when $d$ is 
smaller than the penetration depth. Measurements of the film sheet resistance 
indicate a penetration depth of about 100 nm at $50 \textrm{ cm}^{-1}$, which 
is smaller than the thickness by about a factor of 3. If we assume 
the reflectance is characteristic of a bulk metal, i.e. thick film, plausible 
fits to the data can not be achieved. Still, the film's transmission is less 
than 1\%, so contributions from multiple internal reflections in the substrate 
are negligible and not included in this expression. The solid line is a 
calculation for ${\EuScript R}_S/{\EuScript R}_N$ using the expressions of 
Zimmermann {\it et al.} \cite{Zimmermann} for the optical conductivity of a 
BCS superconductor including the effects of finite scattering. It assumes a 
single gap $2 \Delta = 43 \textrm{ cm}^{-1}$. The solid line in 
Fig. \ref{fig1} assumes a simple model for two gaps, 
$2 \Delta_\pi = 30 \textrm{ cm}^{-1}$ and 
$2 \Delta_\sigma = 56 \textrm{ cm}^{-1}$. 
Here, the optical conductivity is calculated using 
$\hat{\sigma} = f \hat{\sigma}_\pi + (1-f) \hat{\sigma}_\sigma$ where
$\hat{\sigma}_\pi$ and $\hat{\sigma}_\sigma$ are the optical (complex) 
conductivity for a superconductor having energy gaps $2 \Delta_\pi$ and 
$2 \Delta_\sigma$, and $f$ is the spectral weight of the $\pi$ (3D) band. 
This picture is valid for two non-interacting sets of carriers 
responding in parallel. Drude fits to the normal state yielded a scattering 
rate of about $300 \textrm{ cm}^{-1}$. Although the 2 gap model better 
describes the data at low frequencies, the improvement is not sufficient 
to conclude that two gaps must exist.

\begin{figure}
  \begin{center}
    \includegraphics[width=8cm]{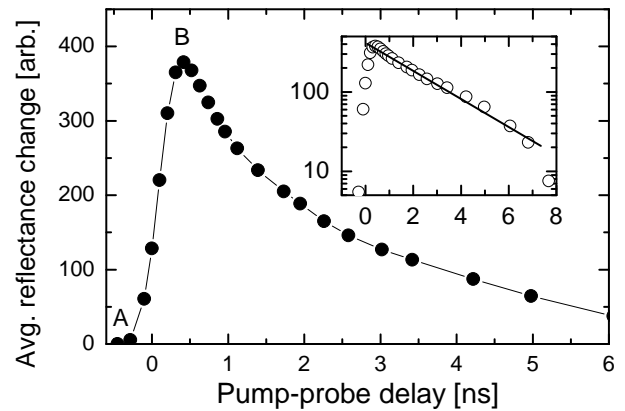} 
  \end{center}
\caption{Spectrally averaged change in the film reflectance as a function 
of time due to a short laser pulse. The average laser power used was 20 mW 
corresponding to 0.4 nJ per pulse. Two points (A and B) mark the moment 
just prior to the pulse (no excess quasiparticles) and at the peak signal 
(end of laser pulse, maximum excess quasiparticle density). The inset 
shows the same data in a semi-log plot. Within experimental error only 
a single exponential decay can be seen in the data.} 
\label{fig2}
\end{figure}

Transient far-infrared photo-reflectance measurements were also performed 
using the same spectrometer and beamline. Here, we measured the reflectance 
change due to illuminating the MgB$_2$ film with 2 ps near-infrared 
($\lambda = 760$ nm) pulses from a Ti:sapphire laser. The laser pulses break 
pairs and weaken the superconducting state for a brief amount of time 
($\approx 1$ ns) \cite{Johnson}. The resulting change in reflectance is 
sensed with the $\approx 1$ ns infrared pulses from the synchrotron in a 
pump-probe configuration. The average laser power was 20 to 50 mW 
(0.4 to 1 nJ per pulse) in a spot filling the whole 3 mm sample aperture.
The pulse repetition frequency was 53 MHz, matching the pulsed 
IR output from the synchrotron. For time-dependent studies, the broadband 
probe pulses were not spectrally resolved, and the response is an average 
of the reflectance across the 10 to $100 \textrm{ cm}^{-1}$ spectral range. 
An analysis of the time dependent transient photo-reflectance gives us a 
good estimate of the effective pair recombination rate \cite{Carr}. 
Figure \ref{fig2} shows the average reflectance change as a function of the 
delay between pump and probe pulses at 6 K. In accordance with other BCS 
superconductors \cite{Johnson,Carr} the effective recombination time is 
found to be a few nanoseconds. For comparison, high-$T_c$ materials have a 
much faster dynamics, lying in the picosecond range \cite{Kaindl2}. One 
important remark is that, within the experimental time resolution, no 
evidence of multiple decays is found. In fact, ultra-fast pump-probe 
measurements on MgB2 \cite{Xu,Demsar1,Demsar2} do not find any evidence 
for more than one relaxation time related to superconductivity down to the 
ps regime. 

\begin{figure}
  \begin{center}
    \includegraphics[width=8cm]{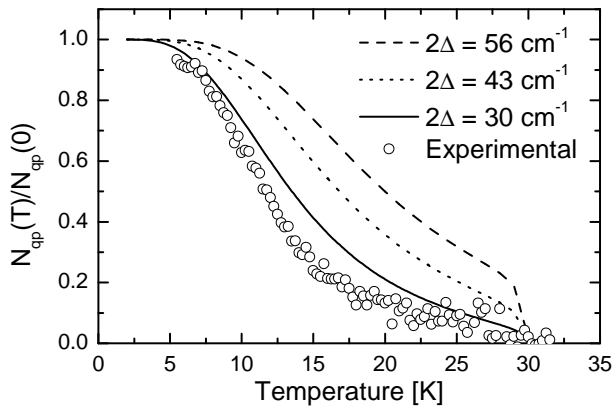} 
  \end{center}
\caption{Temperature dependence of the peak photo-induced reflectance signal 
of Fig. \ref{fig2} (point B). The signal is approximately proportional to 
the number of excess quasiparticles. Also shown are calculations of the excess 
quasiparticle fraction determined using the following parameters (adopting 
\cite{Kaplan} notation): $N(0) = 24 x 10^{21} \textrm{ eV}^{-1}$; 
$N = 11 x 10^{22} \textrm{ cm}^{-3}$; $10^3 b = 0.044 \textrm{ meV}^{-2}$; 
and $<\alpha^2> = 0.56 \textrm{ meV}$. For $2 \Delta_\pi = 30\textrm{ cm}^{-1}$ 
we used $Z_1(0) = 1.45$; for $2 \Delta_\sigma = 56 \textrm{ cm}^{-1}$ we used 
$Z_1(0) = 2.1$ and for $2 \Delta = 43 \textrm{ cm}^{-1}$ we used the average 
value $Z_1(0) = 1.78$. We chose to keep the same gaps used in the equilibrium 
calculation but the data description can be improved if one uses 
$25 \textrm{ cm}^{-1}$ instead of $30 \textrm{ cm}^{-1}$.} 
\label{fig3}
\end{figure}

The amplitude of the time resolved signal is proportional to the number of 
excess quasiparticles in the system $N_{qp}$. Its thermal dependence  
is shown in Fig. \ref{fig3}. Assuming a single gap in the system, 
we can calculate $N_{qp}(T)$ from energy conservation considerations obtaining 
$N_{qp}(T)/N_{qp}(0) = [\Delta(0)/ \Delta(T)]/(1+2 \tau_B/ \tau_R)$ 
\cite{Carr}. We utilized the expressions from Kaplan {\it et al.} \cite{Kaplan}
to calculate the pair recombination time $\tau_R$ and the pair breaking time 
by phonons $\tau_B$. The parameters we used in Kaplan's formalism were 
determined from ref. \cite{Liu}. This single gap picture describes very 
accurately the data as far as we use the small gap value obtained in the two 
component fit for the reflectivity. For comparison, we also show in this figure 
the calculations using the average and the highest $\sigma$ gaps.

\begin{figure}
  \begin{center}
    \includegraphics[width=8cm]{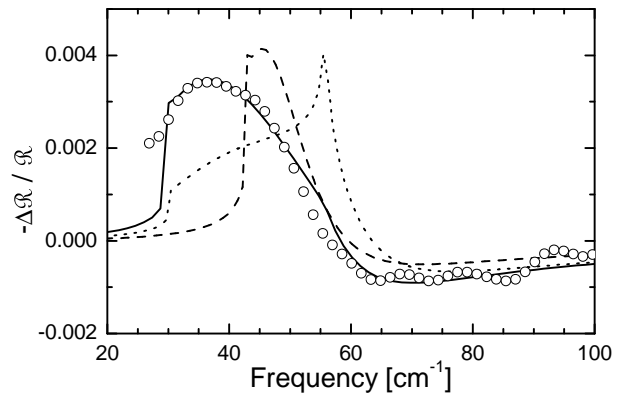} 
  \end{center}
\caption{Photo-induced change in reflectance due to pair breaking by a 1 nJ 
laser pulse (circles). This spectrum is measured between points A and B in 
Fig. \ref{fig2}. The lines are simulations assuming that the photo-induced 
state can be calculated by the difference between two BCS reflectivity 
spectra with slightly different gaps. All the parameters used in the fit 
of Fig. \ref{fig1} are kept here. The dashed line assumes the single average 
gap picture and uses a gap shift of  $0.18 \textrm{ cm}^{-1}$ in the 
photoexcited state. The dotted line is the simulation for a two gap system 
where both gaps shrink in the photoexcited state. We used 
$-0.2\textrm{ cm}^{-1}$ and $-0.38\textrm{ cm}^{-1}$ for the shifts in 
$2 \Delta_\pi$ and $2 \Delta_\sigma$, respectively (corresponding to a 
0.7 \% gap shift in each electron system). The solid curve also uses two 
gaps but assumes that only the $\pi$ gap shifts (by $-0.65 \textrm{ cm}^{-1}$) 
in the photoexcited state.} 
\label{fig4}
\end{figure}

The gap value used in the thermal dependence of $N_{qp}$ is in direct 
contradiction with the equilibrium reflectivity parameters. In the latter, 
regardless of the model used, a gap value around $50 \textrm{ cm}^{-1}$ is 
present whereas only a single gap at $30 \textrm{ cm}^{-1}$ describes 
$N_{qp}(T)$. To address this problem we looked into the photo-reflectance 
spectra, shown in Fig. \ref{fig4}. Measurements were made with the laser 
pulses and synchrotron pulses in coincidence and again with the pulses 
17 ns away from coincidence. From these we calculated the transient 
photo-reflectance as 
$-\delta {\EuScript R}/{\EuScript R} = 
  -({\EuScript R}_{0\textrm{ns}}-{\EuScript R}_{17\textrm{ns}})/
  {\EuScript R}_{17\textrm{ns}}$, 
which essentially eliminates any overall thermal effects from laser heating.
The degree of laser heating could be estimated by measuring the reflectance 
change between laser on and off, and compared with measurements of the actual 
temperature-dependent reflectance. The average temperature rise was less 
than 1 K. A detailed description of non-equilibrium spectroscopy using our 
pulsed laser and synchrotron set up can be found in Ref. \cite{Lobo}. 

The direct consequence of pairs broken by a laser pulse with the creation 
of excess quasiparticles is a weakening of the superconducting state 
\cite{Owen}. This weakened state can be spectroscopically detected and 
resolved as a slightly reduced energy gap \cite{Carr} allowing the transient 
photo-reflectance data to be analyzed using the same expressions as those of 
the equilibrium reflectance. The dashed curve in Fig. \ref{fig4} assumes that 
the system has a single gap at $43 \textrm{ cm}^{-1}$ which is decreased by 
photo-excitation. Although the amplitude and overall shape of the signal can 
be reproduced by this simulation, quantitative agreement is not achieved. 
The dotted line assumes that the system has the same two gaps used in the 
equilibrium reflectance fit and that in the photo-excited state both gaps 
shrink by an amount consistent with a small raise in the electronic 
temperature. The introduction of this second gap does not improve the data 
description and, actually, introduces new features absent from the data. 
Our third approach, depicted by the solid line, assumes that the system 
does have two gaps but that only the smaller energy gap shrinks in the 
photo-excited state. This is the behavior one would expect if, after 
pair-breaking by the laser, the excess quasiparticles are left primarily 
in the system having the smaller energy gap, i.e., quasiparticles are 
``transferred'' from the $\sigma$ band (larger gap) to the $\pi$ band 
(smaller gap) before the recombination is completed. Such a picture 
reconciles completely the whole set of data, equilibrium and non equilibrium.

We can imagine 
two qualitatively different processes that allow for the recombination to 
occur within the lowest gap band only: (i) a transfer of the actual 
quasiparticles (scattering) or (ii) a transfer of energy from the $\sigma$ 
to the $\pi$ band. In the following we discuss such processes. 

When talking about transfer of quasiparticles one could argue for impurity 
scattering diffusing quasiparticles between the bands. The $\pi$ band, 
having a smaller gap is energetically favorable and quasiparticles settling 
to its gap edge would be incapable of returning to the $\sigma$ band. 
This process, however, implies a very large connection between the bands. 
In this case, MgB$_2$ should show only an average gap in the 
equilibrium measurements in disagreement with STM data, for instance 
\cite{Giubileo}.

An alternative is to think in terms of energy transfer instead of quasiparticle 
transfer between the bands. Photo-exciting such a system with a near-IR laser 
pulse will create quasiparticle excitations in both bands. 
As before, these high energy excitations relax by creating lower energy 
quasiparticles and phonons. Looking at the Fermi surfaces, we note that whereas 
the $\pi$ band spans most of the Brillouin zone, the $\sigma$ bands are 
confined to cylinders around the $\Gamma-A$ direction. Therefore, only umklapp 
processes and phonons with momentum along the $\Gamma-A$ direction will be able 
to break other pairs in the $\sigma$ band. We can safely assume that phonons 
created in the relaxation to the gap edge will most likely generate unpaired 
quasiparticles in the $\pi$ band. In addition, quasiparticles that relax to the
$\sigma$ band gap edge will recombine and create $2\Delta_\sigma$ phonons that, 
once again can readily break pairs in the $\pi$ band. The resulting 
quasiparticles can scatter and settle to an energy $\Delta_\pi$ and recombine. 
In addition to the restricted Brillouin zone access to the $\sigma$ band, phonons 
produced by recombination in the $\pi$ band lack energy to break pairs in the 
$\sigma$ band. All in all, both intraband relaxation and quasiparticle 
recombination in the $\sigma$ band contribute to increase the $\pi$ population but 
the opposite process does not happen. Therefore the phonon bottle-neck to 
recombination only occurs for the smallest gap in the system, and the excess 
quasiparticles quickly settle to this gap edge while the quasiparticle density 
in the larger gap band returns to its equilibrium value. The very presence of two 
different energy gaps implies that an excess quasiparticle density in one 
system will preferentially weaken that system over the other. Thus we expect 
that only the smallest energy gap in the system will experience the gap 
reduction described by Owen and Scalapino \cite{Owen}. 

The absence of a two component decay in the time dependent photo-induced signal 
can be explained by the small amount of quasiparticles that recombine in the
$\sigma$ band. A cascading effect makes that each quasiparticle excited by the 
laser photons (energy $h\nu$) creates $h \nu /2 \Delta$ quasiparticles at the 
gap edge. The ``energy transfer'' process proposes that the quasiparticles 
created by the cascading appear at the $\pi$ band. This gives 
us 500 times more quasiparticles in the $\pi$ than in the $\sigma$ band at the 
beginning of the recombination and make it very unlikely to see a fast 
relaxation due to the $\sigma$ band pair recombination.

Recent inelastic x-ray scattering data \cite{Shukla} shows that the $E_{2g}$ 
phonon is anomalously broadened along the $\Gamma-A$ direction in the Brillouin 
zone, the same direction of the $\sigma$ bands. The origin of this broadening 
is related to a strong coupling between this phonon and $\sigma$ electrons
\cite{Joas}. Recent calculations show the importance of band coupling through
phonons \cite{Ord,Lagos} and we propose the $E_2g$ phonon to be the natural 
candidate for the mediator of the energy transfer processes described above.

In this letter we showed static and photo-induced far infrared measurements 
on a MgB$_2$ film. Whereas the equilibrium spectrum indicate a superconducting 
gap around $50 \textrm{ cm}^{-1}$, the non equilibrium dynamics is dominated 
by a gap at $30 \textrm{ cm}^{-1}$. The two sets of data are compatible if 
one considers the picture where excited quasiparticles created by the laser 
impulse relax to the lowest gap band before recombining into Cooper pairs. 
This can be achieved by phonon scattering considering that the $\pi$ band 
covers a much larger volume of the Brillouin zone than the $\sigma$ band. 
We propose to assign the $E_{2g}$ phonon as the mediator of this process.
 
\begin{acknowledgments}
This work was performed with the support of the U.S. Department of Energy 
through contract DE-ACO2-98CH10886 at the NSLS. Operation of the 
synchronized laser system is also supported by DOE through contract 
DE-FG02-02ER45984 with the University of Florida. We are grateful to 
P.B. Allen, J. Carbotte, T. Devereaux, M.V. Klein, I.I. Mazin, A.J. Millis, 
P. Monod, M.R. Norman, E. Nicol and D.B. Tanner, for useful discussions. 
\end{acknowledgments}

\end{document}